\documentclass{aa}  

\usepackage{graphicx}
\usepackage{txfonts}
\usepackage{siunitx}
\usepackage{lineno}
\usepackage{natbib}
\bibpunct{(}{)}{;}{a}{}{,}
\begin{document} 
\title{Advanced modelling of Night Sky Background light for Imaging Atmospheric Cherenkov Telescopes}

\author{
        Gerrit Roellinghoff \inst{1},
        Samuel T. Spencer \inst{1,2}
        and Stefan Funk \inst{1}
       }
\authorrunning{G. Roellinghoff et al.}
\titlerunning{NSB in H.E.S.S.}
\institute{Friedrich-Alexander-Universit{\"a}t Erlangen-N{\"u}rnberg, Erlangen Centre for Astroparticle Physics, Nikolaus-Fiebiger-Str. 2, D 91058 Erlangen, Germany \and Department of Physics, Clarendon Laboratory, Parks Road, Oxford, OX1 3PU, United Kingdom }

\date{Received 14/03/2025 ; accepted 15/05/2025}

\abstract
{}
{A significant source of noise for Imaging Atmospheric Cherenkov Telescopes (IACTs), which are designed to measure air showers caused by astrophysical gamma rays, is optical light emitted from the night sky. This Night Sky Background (NSB) influences IACT operating times and their sensitivity. Thus, for scheduling observations and instrument simulation, an accurate estimate of the NSB is important.}
{A physics-driven approach to simulating wavelength-dependent, per-photomultiplier-pixel NSB was developed. It includes contributions from scattered moonlight, starlight, diffuse galactic light, zodiacal light, and airglow emission. It also accounts for the absorption and scattering of optical light in the atmosphere and telescope-specific factors such as mirror reflectivity, photon detection efficiency, and focal length. The simulated results are corrected for pointing inaccuracies and individual pixel sensitivities and compared to data from the High Energy Stereoscopic System (H.E.S.S.) IACT array. The software package developed for this analysis will be made publicly available.}
{Validation against H.E.S.S.\ data shows small deviations from the prediction, attributable to airglow and atmospheric variability. Per-Pixel predictions provide a good match to the data, with the relative 90\% error range being [-21\%, 19\%]. Compared to the existing standard modelling approach of assuming a constant background, where the relative 90\% error range was [-64\%, 48\%], this is a significant improvement.}
{}

\maketitle
\section{Introduction}
Imaging Atmospheric Cherenkov Telescopes (IACTs) are indirect detection instruments for the observation of very-high-energy gamma rays (>\SI{20}{GeV}). These gamma rays induce particle air showers in the Earth's atmosphere, whose charged component particles cause the emission of Cherenkov light due to their superluminal passage through air. IACTs observe this light by using large mirror arrays to focus it on the focal plane of high-speed cameras, whose pixels consist of photomultiplier tubes (PMTs) or silicon photomultipliers (SiPMs). If a sufficient number of pixels within a time window detect light reminiscent of the shape expected for an air shower, a trigger logic is activated, and all pixel values are read out. The time resolution of IACTs is typically on the order of a few nanoseconds, compared to the arrival time frame of air shower photons of about \SI{10}{\nano\s} \citep{ashton_nectar-based_2020}. This allows for the detection of the faint amount of photons from air showers, which would generally be overshadowed by other sources of photons in the night sky.

The H.E.S.S detector array, stationed \SI{1800}{m} above sea level, consists of two different types of IACTs: an initial four telescopes arranged in a square (CT1-4,  construction finished in 2004), with about \SI{107}{m^2} mirror area each, and a central single dish with about \SI{600}{m^2} mirror area (CT5, constructed in 2012). The original cameras for CT1-4 were upgraded between 2015 and 2017 with NECTAr-based electronics \citep{giavitto_performance_2017, ashton_nectar-based_2020}. The CT5 camera was changed in 2019 to a camera following the FlashCam design developed for future use in the medium-sized telescopes of the Cherenkov Telescope Array Observatory (CTAO) \citep{werner_performance_2017,puhlhofer_science_2022, bi_performance_2022}. Table \ref{table:telescope} shows the properties of CT1-4. The readout of a single pixel is always defined in relation to a pedestal value. This pixel pedestal value depends on the electronic readout noise, gain and the illumination level of the pixel from constant light sources, with higher constant illumination leading to a wider distribution of pedestal values.

Exposure to the night sky results in IACT images containing additional light called Night Sky Background (NSB). Major contributions to NSB are airglow in the upper atmosphere, scattered moonlight, zodiacal light and starlight. Other factors, such as the diffuse galactic light, light pollution and ground reflections, have minor contributions at dark sites. NSB also heavily depends on atmospheric conditions, which in turn depend on telescope location, time of observation and pointing direction. Both the scattering and the absorption of light are dependent on local atmospheric conditions such as aerosol concentration, aerosol albedo, humidity and temperature \citep{garstang_dust_1991,leinert_1997_1998, noll_atmospheric_2012}. 

For valid computational reasons, NSB modelling to date for IACTs has been limited. While there are tools to utilise NSB maps and modelling starlight in IACTs \citep{bernlohr_simulation_2008, buchele_nsb_2020}, the common approach is to use an NSB spectrum taken of the night sky over La Palma by \citet{benn_brightness_1998} \citep{bernlohr_simulation_2008}. The NSB is then scaled based on the expected rate and applied to the entire telescope, with the expected rate being informed by past observations. As such, the NSB is implicitly considered homogeneous over the FoV of the telescope in these cases. However, some areas of the sky observed by IACTs have very inhomogeneous NSB (see Sect. \ref{ch:comparison}). Improved NSB modelling has a wide variety of applications for IACTs, which we discuss in Sect. \ref{ch:usecases}.

With the inclusion of observations under moonlight conditions in the schedules of major IACTs \citep{griffin_veritas_2015, ahnen_performance_2017, tomankova_gain_2022}, reliable NSB predictions have become more important, since some observation positions are rendered impossible because of their proximity to the Moon (as over-exposing PMTs can lead to damage). In the H.E.S.S.\ collaboration, the expected NSB contribution of the Moon was included in the autoscheduler (2020, priv. comm.), using an adjusted model of \cite{krisciunas_model_1991}, which was fitted to photometric data of the sky obtained from Mauna Kea and two sites in Russia in the 1990s. While this includes an estimate for the moonlight and airglow contributions to NSB, it neither deals with the inhomogeneity of NSB over the field of view nor the addition of other NSB sources such as starlight. This can lead to the underestimation of the expected NSB, since starlight-bright regions of the sky (especially our own galaxy) can be comparable to the contribution of moonlight, depending on the separation angle from the Moon. A previous study by \cite{spencer_advanced_2023} on per-pixel NSB in the Small Size Telescopes (SSTs) for CTA included the contribution of starlight, adding V-Band photometric data from the Hipparcos/Tycho and Gaia catalogues to the NSB model from \citet{krisciunas_model_1991}. For this, it uses a modified version of the nsb\footnote{https://pypi.org/project/nsb/} package developed by \citet{buchele_nsb_2020} and colleagues for use by the H.E.S.S.\ collaboration. To the best of our knowledge, there have been no studies comparing simulated NSB to actual NSB values, on a per-pixel level, in the realm of IACTs to date. 

In IACT event reconstruction, gamma-ray air shower parameters are extracted from pedestal-subtracted pixel values. As elevated NSB values increase the pedestal distribution width, it statistically smears out the signal values in each pixel. While this does not introduce energy bias, it worsens energy resolution and the gamma-ray PSF of the instrument. It can also have an effect on the effective area of the instrument, which, depending on the analysis, can have a larger effect \citep{holler_run-wise_2020}. This effect is unique to each NSB pattern, and can not be simulated using homogeneous NSB. IACT arrays currently take observations in so-called `runs' targeting a particular position on the sky, over which observing conditions are assumed to be approximately constant. For H.E.S.S., runs can be up to 28 minutes in duration. Computing approaches such as run-wise simulation incorporate the effect of distinct NSB patterns for each run on event reconstruction, relying on per-pixel NSB values calculated from data. This approach has been shown to match data well \citep{holler_run-wise_2020}.

NSB and its contributing factors have been extensively studied outside of gamma ray astronomy, for examples, see \cite{benn_brightness_1998} for La Palma or \cite{patat_dancing_2008} for Cerro Paranal. \cite{leinert_1997_1998} gives an overview of the measurements and models for the different contributions to the NSB, ranging from ultraviolet to infrared wavelengths. It includes airglow, zodiacal light, diffuse galactic \& extragalactic light and integrated starlight. \cite{noll_atmospheric_2012} introduced and validated an extensive spectrographic model for the NSB at Cerro Paranal in optical and infrared wavelengths. It includes a detailed discussion about atmospheric absorption \& scattering, and was updated with an improved model of lunar contribution by \cite{jones_advanced_2013}. In a similar vein, \cite{masana_multi-band_2021} created GAMBONS, which models the NSB over the sky hemisphere for the purpose of light pollution measurements. Transferring some of the techniques and measurements from other areas of astronomy is therefore an obvious way to improve NSB modelling for IACTs.

The remainder of this work is structured as follows: In Sect. 2, we introduce a new, physically motivated model for the prediction of NSB in IACTs and its components. In Sect. 3 we describe the application of this model to the H.E.S.S.\ telescopes and evaluate its accuracy using actual NSB data covering one year of observations. In Sect. 4 we discuss the validity of the approach, systematics and possible areas of application. Finally, in Sect. 5 we summarise our findings, discuss their implications for further IACT research, and outline further possible areas of improvement for modelling NSB in IACT observations.
\section{Methods}
In general, NSB can be described as the sum of all contributing radiances at an observer location $\Vec{x}$ detected in direction $\Vec{\omega}$ at a specific wavelength $\lambda$
\begin{align}
    I(\lambda, \Vec{\omega}, \Vec{x}) = \sum_i^N I_i(\lambda, \Vec{\omega}, \Vec{x}).
\end{align}
This can be separated into the out-of-atmosphere radiance of $I^0_{i}(\lambda, \Vec{\omega})$ and the atmospheric propagation term $T(\lambda, \Vec{\omega}, \Vec{x})$. The atmospheric propagation term can be further separated into:
\begin{itemize}
    \item Atmospheric Attenuation $T_a(\lambda, \Vec{\omega}, \Vec{x})$, with light in the line-of-sight (LoS) being scattered out of the LoS or absorbed by the atmosphere.
    \item Atmospheric Enhancement $T_s(\lambda, \Vec{\omega}, \Vec{\omega_s}, \Vec{x})$ due to in-scattering, where light from a position $\Vec{\omega_s}$ outside the LoS gets scattered into the LoS.
\end{itemize}
With this, we can write the expected NSB over all sources $i$ of radiance as
\begin{equation}
    \begin{aligned}
        I(\lambda, \Vec{\omega}, \Vec{x}) = & \sum_i^N T_a(\lambda, \Vec{\omega}, \Vec{x})I^0_{i}(\lambda,\Vec{\omega})\\ & + \sum_i^N \int_{\Omega_s} T_s(\lambda, \Vec{\omega}, \Vec{\omega_s}, \Vec{x})I^0_{i}(\lambda, \Vec{\omega_s})\,d\Vec{\omega_s}.
    \end{aligned}
\end{equation}
Finally, the photon flux $\Phi(\Vec{\omega}, \Vec{x})$ depends on the bandpass $S(\lambda)$ of the instrument measuring the NSB:
\begin{equation}\label{eq:intensity}
    \begin{aligned}
        \Phi(\Vec{\omega}, \Vec{x}) = & \int_0^{\infty} d\lambda S(\lambda) \left(\sum_i^N T_a(\lambda, \Vec{\omega}, \Vec{x})I^0_{i}(\lambda,\Vec{\omega})\right. \\ & + \left. \sum_i^N \int_{\Omega_s} T_s(\lambda, \Vec{\omega}, \Vec{\omega_s}, \Vec{x})I^0_{i}(\lambda, \Vec{\omega_s}) d\Vec{\omega_s} \right).
    \end{aligned}
\end{equation}
For diffuse sources, the integral in the second term is often approximated as a modification to the attenuation term \citep{kwon_observational_2004,kocifaj_kocifaj_2009,noll_atmospheric_2012,winkler_revised_2022}, resulting in an effective atmospheric transmittance $T_{\mathrm{eff}}(\lambda, \Vec{\omega}, \Vec{x})$:
\begin{equation}\label{eq:intensity_eff}
    \begin{aligned}
        \Phi(\Vec{\omega}, \Vec{x}) = & \int_0^{\infty} d\lambda S(\lambda) \sum_i^N T_{\mathrm{eff}}(\lambda, \Vec{\omega}, \Vec{x})I^0_{i}(\lambda,\Vec{\omega}).
    \end{aligned}
\end{equation}
In general, the above equations also depend on the time $t$, especially in the context of coordinate transforms. To transform to source dependent reference frames, we need to know the transformation from local coordinates $(\Vec{\omega}, \Vec{x})$ to galactic $(l,b)$, equatorial $(\alpha, \delta)$ or heliocentric ecliptic $(\Lambda, \beta)$ coordinates. For this we use the AstroPy\footnote{http://www.astropy.org} software package \citep{astropy_collaboration_astropy_2013, astropy_collaboration_astropy_2018, astropy_collaboration_astropy_2022}.

To calculate the photon rate for a single pixel in a telescope, the local coordinates $(\Vec{\omega}, \Vec{x})$ need to be transformed to telescope coordinates $(x,y)$ depending on the direction $\Vec{\eta}$ the telescope is pointing. The total rate $F_{\mathrm{SIM}}$ for pixel $j$ is then equivalent to
\begin{equation}\label{eq:pixelrate}
    F_{\mathrm{SIM}}(\Vec{\eta}) =  \int_{A_j} dxdy \int_{\Omega} d\omega P(x, y | \Vec{\omega}, \Vec{\eta}) \Phi(\Vec{\omega}, \Vec{x}),
\end{equation}
with telescope point source function $P(x, y |\Vec{\omega})$, pixel area $A_j$ and telescope pointing direction $\Vec{\eta}$. Computationally, this can be approximated as
\begin{equation}
    F_{\mathrm{SIM}}(\Vec{\eta}) =  \sum_{(x, y) \in A_j} w_{(x,y)} \sum_{\Vec{{\omega}} \in C_j(\Vec{\eta})}  P(x, y | \Vec{\omega}, \Vec{\eta}) \Phi(\Vec{\omega}, \Vec{x}) 
\end{equation}
with $w_{x,y,}$ the weight assigned to position $(x,y)$ on the pixel for the corresponding discrete integration scheme, and $C_j(\Vec{\eta})$ being the set of positions in the sky with significant contribution to the rate in pixel $j$. 
\subsection{Sources of NSB}

\subsubsection{Moonlight}\label{ch:moonlight}
In the context of NSB, the most significant source in the night sky is the Moon. During the brightest lunar phase of full moon, with an average visual magnitude of $-12.7$, its radiative flux is about 2000 times higher than the second brightest object (Venus). As such, as long as the Moon is above the horizon and its phase angle is small (Table 2, \citet{krisciunas_model_1991}), moonlight is the dominating contribution to the NSB at most positions on the sky. 

\cite{krisciunas_model_1991} used an empirical fit of the scattering function to data from Mauna Kea and two Russian sites and an empirical function modifying the moon brightness based on lunar phase to create a formula for the expected lunar contribution in the V band. Being one of the first formulas for moonlight brightness, they achieved an rms uncertainty for the V-band of 23\%. \cite{jones_advanced_2013} improved on this concept by incorporating an accurate solar spectrum, a lunar albedo fit to data by \cite{kieffer_spectral_2005} and a physically based atmospheric model. They reached an uncertainty of less than 20\% over the optical range when comparing their spectra to measured FORS1 spectra. \cite{winkler_revised_2022} modified their approach, differing in the use of an analytical treatment of atmospheric scattering and their interpolation of lunar albedo values. When fitting the offset, their resulting uncertainty is typically on the order of less than 5\% in the UBVRI bands. For the light emission component of the moonlight, we follow \citet{jones_advanced_2013}, with our atmospheric scattering approach being described in Sect. \ref{ch:scattering}. 

The intensity $I_{lun}$ of moonlight entering the Earth's atmosphere following \cite{jones_advanced_2013} is related to the solar light intensity $I_{\odot}$ via
\begin{equation}
    I_{lun}(\lambda) = I_{\odot}(\lambda) \frac{\Omega_M}{\pi} \left(\frac{384,400}{D_{M}}\right)^2  A(\lambda).
\end{equation}
with $\Omega_M$ being the solid angle of the Moon ($\Omega_M = \SI{6.4177e-5}{sr}$) and $D_{M}$ the distance to Earth. The solar spectrum is taken from \cite{rieke_absolute_2008}. The lunar Albedo $A$ is a wavelength-dependent fit to observational data performed by \citet{kieffer_spectral_2005}:
\begin{equation}
\begin{aligned}
      ln(A(\lambda)) = & \sum_{i=0}^3 a_{i,\lambda}g^i + \sum_{j=1}^3 b_{j,\lambda}\Phi^{2j-1} \\& + d_{1,\lambda}e^{-g/p_1} + d_{2,\lambda}e^{-g/p_2} + d_{3,\lambda}\cos[(g-p_3)/2]  
\end{aligned}
\end{equation}
with lunar phase parameter $g$ and solar selenographic longitude $\Phi$. The values for $p_n, a_i, b_j, d_n$ are taken from \cite{kieffer_spectral_2005}. Then a quadratic spline is fitted to determine $I_{lun}(\lambda)$ in a smooth manner. For our purposes, the moon is assumed to be a point source due to the fact that H.E.S.S. observations are separated by at least \SI{10}{\degree} from the Moon.
\subsubsection{Zodiacal light}
Zodiacal light is caused by solar light being reflected off interplanetary dust grains in the ecliptic. As such, it is highly dependent on the ecliptic coordinates, with the NSB rate decreasing with increasing ecliptic latitude $|\beta|$. There is also the phenomenon of gegenschein at the anti-solar point ($\lambda - \lambda_{\odot} = 180^{\circ}$), where there is a local increase in zodiacal light \citep{kwon_observational_2004}. 

To model zodiacal light, we rely on observational \SI{500}{nm} data taken by \cite{dumont_zodiacal_1976}, as summarised in Table 16 in \cite{leinert_1997_1998}. To get values for all ecliptic coordinates ($\Lambda$, $\beta$), a 2D linear interpolation is used to obtain $f_{zl}(\Lambda, \beta)$. For the ecliptic pole $\beta = 90^\circ$, the brightness is fixed to $\SI{60}{S_{10}}$. For the Spectral Energy Distribution (SED) we assume the solar spectrum from \cite{rieke_absolute_2008}, as in (Sect. \ref{ch:moonlight}). We further use the colour correction coefficient $f_{co}$ from \cite{leinert_1997_1998} to account for reddening of the zodiacal light in relation to the solar spectrum:
\begin{align}
    I_{zl} &= f_{co}(\lambda)f_{zl}(\Lambda, \beta)I_{\odot}(\lambda).
\end{align}
\subsubsection{Airglow}\label{ch:airglow}
Airglow is a natural phenomenon in Earth's upper atmosphere, where atoms and molecules shed accumulated excess energy. Unlike other sources of NSB, such as starlight or moonlight, it is completely diffuse and emitted from different atmospheric layers, starting at around \SI{87}{\kilo\meter} and reaching up to \SI{250}{\kilo\meter}. It exhibits both a line emission and a continuum component, with a variety of contributions from chemiluminescent and recombination processes. It varies considerably over short (minutes) to long (years) timescales, stemming from a variety of factors, such as solar activity cycle, seasonal changes in atmospheric composition, atmospheric gravity waves and time of night. This inherent variability of the airglow makes predictive modelling challenging, resulting in a mismatch with individual observations (see Sect. \ref{ch:ag_var}). For a more detailed description of airglow and its variability, see \citet{khomich_airglow_2008}.

Since IACT bandwidth generally spans multiple $\SI{100}{nm}$, we add the line emission components into the continuum component. The airglow spectrum is taken from \cite{noll_atmospheric_2012}, generated by the ESO SKYCALC tool \citep{noll_atmospheric_2012,jones_advanced_2013} for Cerro Paranal for a solar flux level of 73 solar flux units (sfu), the average solar flux over the one year of observations used in Sect. \ref{ch:application}. To account for differences in solar flux during the time period of observations, we follow the approach by \cite{noll_atmospheric_2012} and use a scaling factor for each run of
\begin{align}
    f_{sf} &= 0.2 + 0.00614 F_{10.7}
\end{align}
with $F_{10.7}$ being the solar flux level in sfu\footnote{Data from https://www.spaceweather.gc.ca, accessed 2024-12-11.}, to account for enhancement of airglow due to increased solar activity.
The airglow intensity also changes depending on the Zenith angle, due to the increased volume of air within the LoS. This dependence can be approximated following \citet{van_rhijn_brightness_1921} as
\begin{align}
    f_{z}(Z) = \frac{1}{\left(1 - \left(\frac{R}{R+H}\right)^2 \sin^2(Z) \right)},
\end{align}
also known as the van Rhijn effect or formula. The total formula for the airflow, depending on zenith and solar flux, is then
\begin{align}
    I_{ag}(\lambda, Z, F_{10.7}) = f_{sf} f_{z} I_{ag}(\lambda).
\end{align}

\subsubsection{Starlight}\label{ch:starlight}
Depending on the location of the sky, starlight is one of the highest contributors of NSB. For positions close to the galactic centre, it can be the brightest component on moonless nights \citep{masana_multi-band_2021}. It is also the most localised, with the light from bright stars being the dominant source of NSB in pixels they are projected into. In IACT cameras, very bright stars can saturate pixels. In the case of conventional PMT pixels, the high resulting currents can lead to accelerated ageing. To avoid this, such pixels are generally turned off, either pre-emptively or in response to high currents.

With the advent of large surveys, such as Hipparcos-Tycho \citep{esa_hipparcos_1997} and Gaia \citep{gaia_collaboration_gaia_2016, gaia_collaboration_gaia_2023}, the stellar contribution to the NSB can be modelled for each star. Since the GaiaDR3 dataset has issues with completeness in the low-magnitude regime (G<3) due to detector over-saturation \citep{fabricius_gaia_2021}, a combination of the Hipparcos-Tycho and GaiaDR3 catalogues was used to achieve a high degree of completeness in both the low- and high-magnitude regimes. For this, we searched for Hipparcos stars below $H_{mag} <4$ without counterparts in GaiaDR3 using the Simbad database \citep{wenger_simbad_2000}. We then added these stars to our catalogue.

To approximate the SED of stars, we use synthetic spectra from \cite{coelho_new_2014}. To assign spectra to stars, their photometric colours are determined and the closest match from the spectral library is used. It is then scaled to the G-Band magnitude for GaiaDR3 or the H-Band magnitude of Hipparcos-Tycho.

To reduce computational complexity, this is only done for stars with $G_{mag} < 15$. In total there are 36,908,056  stars from the GaiaDR3 catalogue and 93 supplementary stars from the Hipparcos catalogue for $G_{mag} < 15$. Stars with $G_{mag} > 15$ are grouped into HEALPix pixels \citep{gorski_healpix_2005} for reasons of computational efficiency, and their magnitudes $M_i$ are combined and then treated like a single star.

\subsubsection{Diffuse galactic light}\label{ch:galactic}
The Diffuse Galactic Light (DGL) is caused by the scattering of starlight on dust grains in the interstellar medium. It typically contributes between 20\% and 30\% of the integrated total light from the Milky Way \citep{leinert_1997_1998}. Measurements of the DGL in the optical regime have been performed using Pioneer 10 photometry \citep{toller_study_1981}, but did not deliver a comprehensive measurement for the entire galactic plane and were heavily contaminated by stellar light \citep{leinert_1997_1998}. For our purpose, we avoid using DGL measurements directly. Instead, we use an approach following \cite{kawara_ultraviolet_2017}. Using data from the \cite{schlegel_maps_1998} dust maps, we use their observed relation between $\SI{100}{\mu \meter}$ and optical wavelength emission, to model the optical emission as
\begin{align}
    I_{v,i} (DGL) &= b_i I_{v,100} - c_i I^2_{v,100} \\
    I_{v,100} &= I_{v, \mathrm{SFD}} - \SI{0.8}{MJy sr^{-1}} 
\end{align}
The individual values for $b_i$ and $c_i$ are taken from \cite{kawara_ultraviolet_2017} and linearly extrapolated to values outside of the given \SIrange{230}{650}{nm} range.
This model is not valid for optically thick regions, for example, at low galactic latitudes. To avoid unrealistic values, we adopt the approach by \cite{masana_multi-band_2021}; for every position, we impose a maximum value of the DGL of 0.35 times the integrated stellar radiance value of each pixel, inspired by the highest values reported in \cite{toller_study_1981}.
\subsection{Atmospheric scattering and extinction}\label{ch:scattering}
When passing through the Earth's atmosphere, light gets scattered by aerosols and molecules in the optical path. This leads to an extinction in the direction of the source and enhancement from in-scattered light from sources outside of the line of sight. The attenuation of a source of intensity $I^0_s$ on top of the atmosphere and airmass $X$ can be described using the extinction coefficient $\tau$:
\begin{align}
    I_s(\lambda, X) = T_a(\lambda, \Vec{\omega}, \Vec{x}) I^0_s(\lambda)= e^{-\tau (\lambda) X(\Vec{\omega}, \Vec{x})} I^0_s(\lambda).
\end{align}
$\tau$ can be expressed in terms of Rayleigh scattering coefficient $\tau_R$, Mie scattering coefficient $\tau_M$ and molecular absorption coefficient  $\tau_A$:
\begin{align}
    \tau(\lambda) = \tau_R(\lambda) + \tau_M(\lambda) + \tau_A(\lambda).
\end{align}
The Rayleigh scattering coefficient $\tau_R$ can be assumed to be stable and taken from \citet{leckner_spectral_1978} to be
\begin{align}
    \tau_R(\lambda) &= 0.008735 \lambda^{-4.08} \times e^{\frac{H}{\SI{8.47},{km}}},
\end{align}
with $H$ being the height of the observation site. This was chosen for its simple parametrisation. More complex parametrisations exist, but differences over the optical wavelength range are on the sub-percentage level. The Mie scattering coefficient, describing aerosol scattering, changes depending on atmospheric conditions. A wavelength-dependent value can be determined from the angstrom exponent $\alpha$ and scattering coefficient $\tau_{\lambda_0}$ at reference wavelength $\lambda_0$ \citep{leckner_spectral_1978}
\begin{align} \label{eq:aod}
    \tau_M(\lambda) &= \tau_{\lambda_0} \left(\frac{\lambda}{\lambda_0}\right)^{-\alpha},
\end{align}
with $\tau_{\lambda_0}$ and $\lambda_0$ depending on local conditions. A good approximation are values from AERONET sites, with one being the H.E.S.S.\ site itself since 2016. AERONET is a global network of sun-photometers measuring direct and scattered sunlight, and uses these measurements to determine atmospheric aerosol properties \citep{giles_advancements_2019}. However, AERONET measurements are typically taking during the day and H.E.S.S. measurements during the night, as such, the accuracy of using AERONET measurements hinges on the method of interpolation. The molecular absorption coefficient $\tau_A$ is dominated by water vapour, molecular oxygen and ozone in the optical range \citep{bogumil_measurements_2003}. We use the ESO SKYCALC tool \citep{noll_atmospheric_2012,jones_advanced_2013} to calculate the absorption curve for a standard atmosphere with precipitable water vapour being \SI{2.5}{mm}. The choice of precipitable water vapour level is arbitrary, as the water vapour absorption lines are all at wavelengths longer than \SI{650}{nm}, which is the upper end of the H.E.S.S. passband \citep{noll_atmospheric_2012}.

The second term in Eq. \ref{eq:intensity} describes the enhancement of light within the LoS due to in-scattering of light from other sources. As the atmosphere is optically thin, we use a single scattering approach with correction term as in \citet{kocifaj_kocifaj_2009, winkler_revised_2022}, with the in-scattering being described by
\begin{align}
    T_s(\lambda, \Vec{\omega}, \Vec{\omega_s}, \Vec{x}) = p(\lambda, \Vec{\theta}) \Gamma (\lambda, z, z_s)
\end{align}
with indicatrix $p(\lambda, \Vec{\theta})$ and gradation function
\begin{align}
    \Gamma(\lambda, z, z_s) = \frac{e^{-\tau (\lambda) \sec(z)} - e^{-\tau (\lambda) \sec(z_s)}}{\sec(z_s) - \sec(z)}.
\end{align}
with $z$ and $z_s$ being the zenith of the observation direction and the source direction, respectively.
The indicatrix $p(\lambda, \Vec{\theta})$ is the spatial scattering function, depending on the separation angle $\Vec{\theta}$ between $\Vec{\omega_s}$ and $\Vec{\omega}$ and the respective extinction coefficients. Following \cite{kocifaj_kocifaj_2009} it can be described as the combination of Rayleigh and Mie Scattering functions (for high aerosol albedos), weighted by their respective scattering coefficient 
\begin{align}
    p(\lambda, \Vec{\theta}) = \frac{\tau_R(\lambda)}{\tau(\lambda)} p_R(\Vec{\theta}) + \frac{\tau_M(\lambda)}{\tau(\lambda)} p_M(\Vec{\theta}).
\end{align}
Rayleigh scattering (disregarding polarisation) can be described by
\begin{align}
    p_R(\Vec{\theta}) = \frac{1}{4\pi}\frac{3}{4}(1+\cos^2(\theta))
\end{align}
and Mie scattering approximated with the Henyey-Greenstein \citep{henyey_diffuse_1941} function:
\begin{align}
    p_M(\Vec{\theta}) = \frac{1}{4\pi}\frac{1-g^2}{(1+g^2-2g\cos(\theta))^{3/2}},
\end{align}
where $g$ is the asymmetry factor, which is typically between 0.5 and 0.9 for the terrestrial atmosphere, depending on aerosol conditions. For our purposes, we adopt a value of 0.8 as determined to be the best fit value to moonlight data in \cite{winkler_revised_2022}.

As described in Eq.\ref{eq:intensity_eff}, the in-scattering can be approximated by using an effective atmospheric transmittance:
\begin{align}
    I(\lambda, X) = T_{\mathrm{eff}}(\lambda, \Vec{\omega}, \Vec{x}) I_s(\lambda) = e^{-f_{\mathrm{eff}}\tau(\lambda)X(\Vec{\omega}, \Vec{x})}I_s(\lambda)
\end{align}
The effective scattering coefficient $f_{\mathrm{eff}}$ can depend on multiple arguments and differ between sources. Commonly, a fixed argument is used for all diffuse components. In \cite{noll_atmospheric_2012}, $f_{\mathrm{eff}}$ depends on the airmass for airglow and top-of-the-atmosphere intensity for zodiacal light. They based these values on simulations done with their 3D scattering code. For our purposes, we use a factor of $f_{\mathrm{eff}} = 0.75$ for the zodiacal light as described by \cite{kwon_observational_2004}, and a fit to the airmass for the airglow inspired by \cite{noll_atmospheric_2012}:
\begin{align}
    f_{\mathrm{eff}}(X) = 1.7\log(X) - 0.146.
\end{align}
\subsection{Telescope model}\label{ch:telescope}
\begin{table}
\caption{Telescope Parameters for the H.E.S.S. I telescopes}
\label{table:telescope}
\centering
\begin{tabular}{c c c c}
\hline\hline
Parameter & Value \\
\hline
Mirror Dish Area & \SI{107}{m^2} \\
Focal Length& \SI{15}{m} \\
Object focal distance & \SI{10}{km} \\
PMT pixels & 960 \\
PSF (on-axis) & \SI{0.03}{\deg} (RMS) \\
\hline
\end{tabular}
\end{table}
\begin{figure}
  \resizebox{\hsize}{!}{\includegraphics{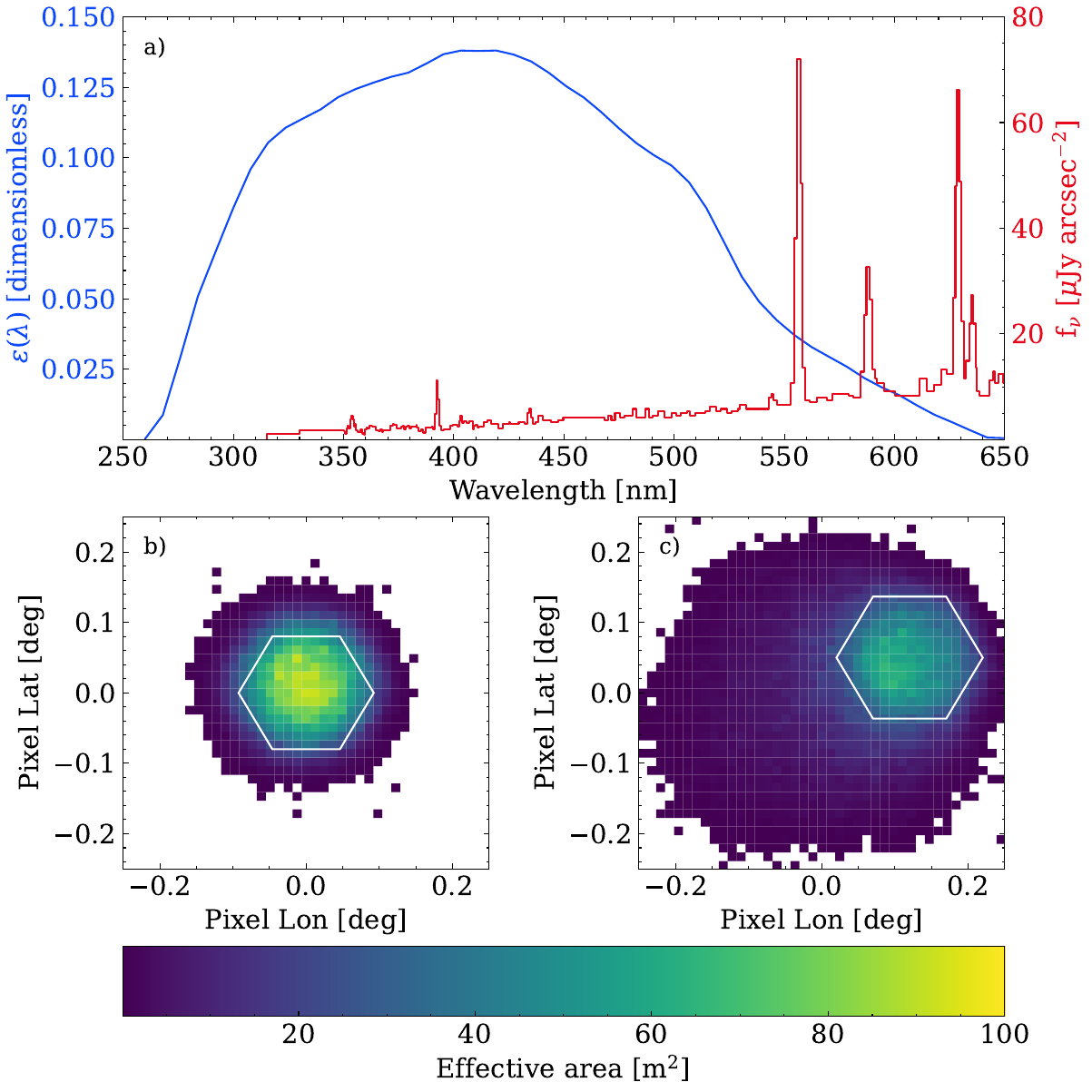}}
  \caption{The assumed instrumental response for the individual H.E.S.S. I telescopes. a) shows the total bandpass, with $\epsilon(\lambda)$ being defined as the ratio between detected and incident photons. For comparison, the night sky spectrum from \cite{benn_brightness_1998} is shown in red. b) shows the effective area for a pixel close to the optical axis. c) shows the effective area for an off-axis pixel. This shows a noticeable asymmetry due to optical aberrations. The white hexagons in b) and c) note the pixel size when projected into the reference frame of the telescope.}
  \label{fig:telescope_model}
\end{figure}
We focus our analysis on the H.E.S.S. CT1 telescope as an exemplar of the "third generation" of IACTs. CT1 follows a Davies-Cotton design, with spherical mirrors arranged on a partial sphere with a radius corresponding to their focal length. In the cases of  CT1, there are 380 mirror facets with a diameter of \SI{60}{cm} each, giving the telescopes an effective aperture of around \SI{107}{m^2}. The camera of the telescope consists of 960 photomultiplier tubes acting as pixels, with Winston cones concentrating the light onto them. Table \ref{table:telescope} shows the specifications of H.E.S.S. CT1-4. With these optical components in the light path, the bandpass of the telescope can be described by 
\begin{align}
    \epsilon(\lambda) &= R(\lambda) E(\lambda) P(\lambda),
\end{align}
with mirror reflectivity $R$, Winston cone efficiency $E$ and photon detection efficiency $P$. While these might vary between individual mirrors, Winston cones and PMTs, in the context of this work, we assume a shared bandpass for all components. Fig. \ref{fig:telescope_model} shows the bandpass for the H.E.S.S. I telescopes, calculated from $R(\lambda)$ and $E(\lambda)$ given in \citet{bernlohr_optical_2003}, and $P(\lambda)$ in \citet{bernlohr_simulation_2008}. To account for differences between PMTs to first order, we perform a quasi-flatfielding procedure outlined in Sect. \ref{ch:flatfield}. Additionally, a factor of 0.8 is applied to account for telescope degradation compared to the nominal design value of the optical efficiency \citep{schafer_simulation_2023}.

The optical point spread function (PSF) is similar to the pixel size of 0.02 square degrees for CT1-4, but suffers from optical aberrations for off-axis light. To determine the response of pixels to light, we utilise ray-tracing to determine the amount of light hitting a pixel depending on the coordinate of the point source in the telescope frame, resulting in an effective area for each coordinate and pixel. Fig. \ref{fig:telescope_model} shows the resulting effective area for a pixel close to the optical axis and at the edge of the telescope, highlighting the effect of the off-axis optical aberrations. New generation IACT designs, such as the Schwarzschild-Couder design for the CTA SST telescopes, suffer less under these off-axis aberrations as a result of to their two-mirror design \citep{giro_first_2017}.

The PSF model does not account for telescope deformation or changes to the PMT response over time (besides the Flatfielding coefficients). But the optical PSF of the H.E.S.S. telescopes has proven to be remarkably stable (2024, priv. comm.), and deformations are small for altitude angles above \SI{40}{\degree} \citep{cornils_optical_2003}.
We have implemented all discussed models in an open-source software package called nsb2\footnote{https://github.com/GerritRo/nsb2}. Depending on the observed star field, simulations run at a speed of around \SI{1}{Hz}.

\section{Application to H.E.S.S.}\label{ch:application}
As the readout windows of the H.E.S.S. telescopes are on the order of nanoseconds, and typical NSB rates are of the order \SIrange{50}{1000}{MHz}, the integrated NSB signal in a single Cherenkov shower event image is too low for a statistical comparison to simulation data. As such, for H.E.S.S. CT1-4, the following procedure is used to estimate the NSB in a pixel:
\begin{enumerate}
    \item For each event image, determine all pixels containing Cherenkov light by checking if it or its neighbours are above a determined threshold.
    \item For each event image, read out the intensity value for each pixel determined not to contain Cherenkov light. This is called the pedestal value.
    \item When there is enough data for each pixel, the root-mean-square (RMS) of these pedestal values is determined.
\end{enumerate}
As the CT1-4 PMTs are AC coupled to their samplers, the mean pedestal value of a CT1 pixel is NSB independent and cannot be used to determine incident NSB. Since higher NSB does result in larger fluctuations in the pedestal distribution, the RMS can be related to the NSB rate by \citep{aharonian_calibration_2004}
\begin{align}
    F_{\mathrm{OBS}} = \frac{1}{\tau}\sqrt{RMS_{\rm p}^2 - RMS_0^2 - \sigma_{\gamma_e}^2}
\end{align}
with $RMS_{\rm p}$ being the pedestal RMS, $RMS_0$ the electronic pedestal in the absence of illumination, $\sigma_{\gamma_e}$ the charge dispersion of a single photoelectron and $\tau$ the readout window. To gain enough pedestal values per pixel to determine the RMS accurately takes around \SI{8}{s} in CT1-4. For CT5, a more accurate method can be used due to its DC coupling, which in principle enables the NSB readout up to event-wise rates, see \citet{werner_performance_2017} For CT1-4, the expected relative error on the resulting NSB rate $F_{\mathrm{OBS}}$ is within \SIrange[]{10}{20}{\%} \citep{aharonian_calibration_2004}. For each pixel, $F_{\mathrm{OBS}}$ can then be directly compared with the simulated NSB rate $F_{\mathrm{SIM}}$ for each readout window.

\subsection{The NSB dataset}
As mentioned in Sect. \ref{ch:telescope}, we focus our analysis on the CT1 telescope. We use NSB data taken during 1 year of CT1 observations for our comparison. The time period is from the 8th of June 2020 until the 3rd of June 2021. This was chosen to coincide with a period of time where extensive Monte-Carlo validation was performed for the H.E.S.S. instrument \citep{leuschner_validating_2023}.

\begin{figure}
  \resizebox{\hsize}{!}{\includegraphics{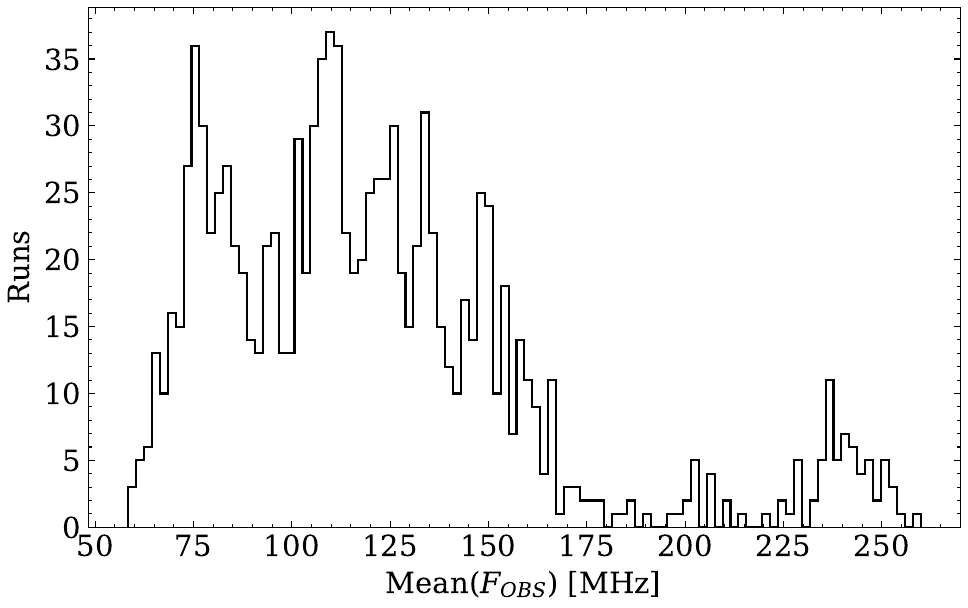}}
  \caption{Distribution of mean $F_{\mathrm{OBS}}$ over all pixels for each run. Galactic fields are found towards the upper end, extragalactic fields are at the lower end.}
  \label{fig:nsb_dist}
\end{figure}

In total, this period contains 2618 observation runs for the CT1 telescope. To remove unwanted effects such as cloud coverage or unstable atmospheric conditions, we imposed quality cuts on this dataset: removing all runs where the telescope event rate fluctuated or drifted\footnote{Where the relative change of the system trigger rate during a run is less than 3\% and the fluctuation of the system trigger rate as a percentage of the mean value is less than 1.5\%.} (indicators for clouds and/or unstable atmospheric conditions). We also removed all runs with a duration lower than \SI{28}{min}, since premature cancellations of runs can be potentially indicative of problems during the run. We further impose a cut on the average altitude of each observation run to be above \SI{40}{\deg}, to stay in a regime where the optical PSF of the telescope is stable \citep{cornils_optical_2003}. After all cuts, there are 1231 Observation runs left, with 97 being observations during partial moonlight; the gain settings for the camera are the same during both modes of operation \citep{tomankova_gain_2022}. The runs include both extragalactic and galactic observation positions and are distributed over the entire area of the sky observable to H.E.S.S.. In total, there are 59 different observational fields present in this dataset. The mean NSB rate $F_{\mathrm{OBS}}$ for runs varies between \SI{60}{MHz} and \SI{260}{MHz} (see Fig. \ref{fig:nsb_dist}).

\subsection{Pointing correction and quasi-flatfielding}\label{ch:flatfield}
\begin{figure}
  \resizebox{\hsize}{!}{\includegraphics{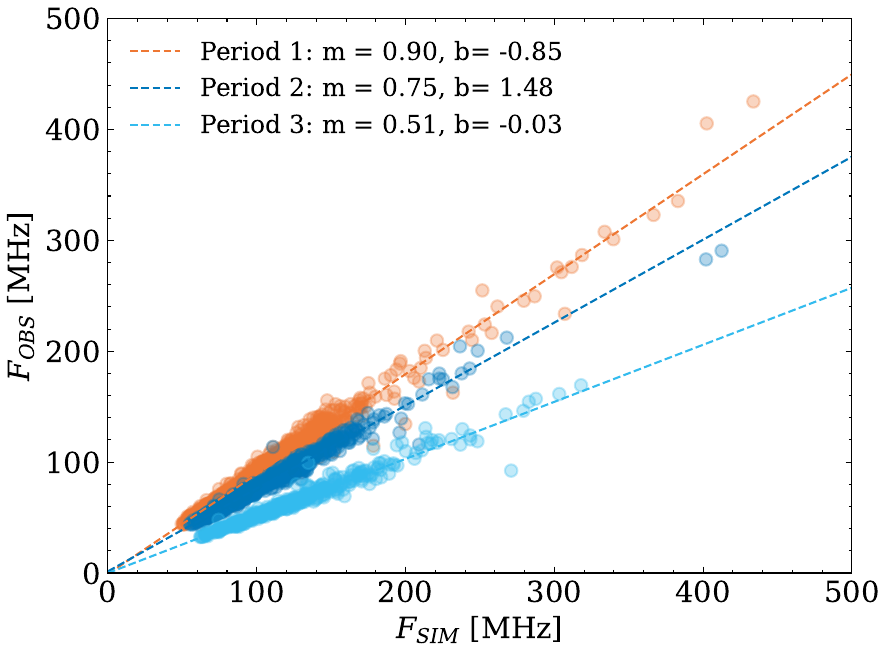}}
  \caption{Determination of flatfielding coefficients for a pixel close to the camera centre showing large changes between time periods. The coefficients are determined for three distinct time periods, to account for gain readjustment at the beginning of the periods. Each scatter point is an NSB datapoint determined during a run. The transition dates between periods are 2020-11-02 and 2021-04-29. This pixel is one of the most affected pixels.}
  \label{fig:flatfield}
\end{figure}
To ensure the best possible match between data and simulations, we corrected the simulated field using the H.E.S.S.\ pointing correction for each time period, which have been estimated to be accurate within \SI{8}{\arcsec} \citep{hinton_status_2004}. These corrections rely on pointing runs taken at a monthly frequency, which centre the telescope on guide stars and note the deviation depending on altitude and zenith. Additionally, a refraction model is used to simulate the refraction of starlight. 

While the H.E.S.S.\ telescopes take flatfielding data using LED flashers, we have found this data to be unable to correct NSB images. This is most likely due to the different way NSB data is captured compared to the way flatfielding for air shower observations by H.E.S.S. is performed (uniform exposure compared to short flashes). As such, we implement a quasi-flatfielding technique: 
\begin{enumerate}
     \item We fit an empirical scaling factor to each model component for each run. This is done to capture the influence of atmospheric conditions and airglow variability. To minimise the contribution of single pixel outliers, a soft\_l1 loss function is used for the fit. This is defined as
\begin{equation}
\mathcal{L}(x) = 2 (\sqrt{1+x^2} -1).
\end{equation}
     \item We then perform a linear fit for each individual pixel, again using a soft\_l1 loss function to remove the influence of single pixel outliers. We apply this fit for three different time periods, necessitated by PMT gain changes during the year affecting pixel response (see Fig. \ref{fig:flatfield}).
\end{enumerate}
For this quasi-flatfielding, we have excluded runs with Eta Carinae in the FoV, since the bright nebulae component is not captured by our model (see Sect. \ref{ch:etacar}) and worsens the fit considerably. The data is then corrected using the determined linear coefficients $(m,b)$ depending on the time period. The fitting results from step 1 can also be used to determine the airglow (section \ref{ch:ag_var}) and atmospheric conditions (Sect. \ref{ch:atm_fit}).
\subsection{Comparison between simulation and NSB data}\label{ch:comparison}
\begin{figure*}
\centering
   \includegraphics[width=17cm]{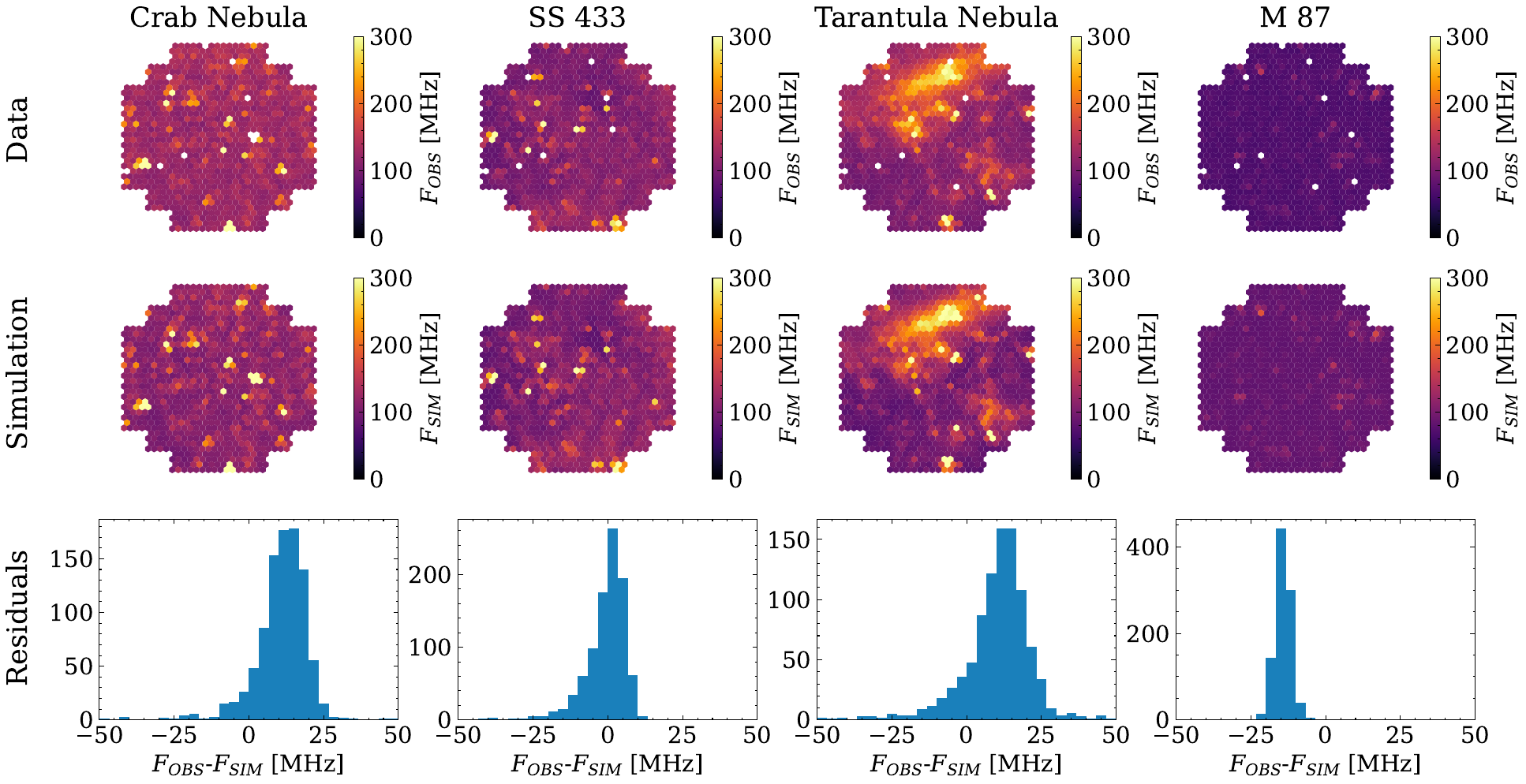}
     \caption{Comparison between NSB measurements and simulation for the Crab Nebula, SS 433, Tarantula Nebula and M 87. The upper row is the data corrected for flatfielding, the middle row the simulated data and the bottom row the pixel-wise residuals in MHz. White pixels in the top row mean that this specific pixel was disabled during the time of data acquisition.}
     \label{fig:individual_comp}
\end{figure*}
\begin{figure*}
\centering
   \includegraphics[width=17cm]{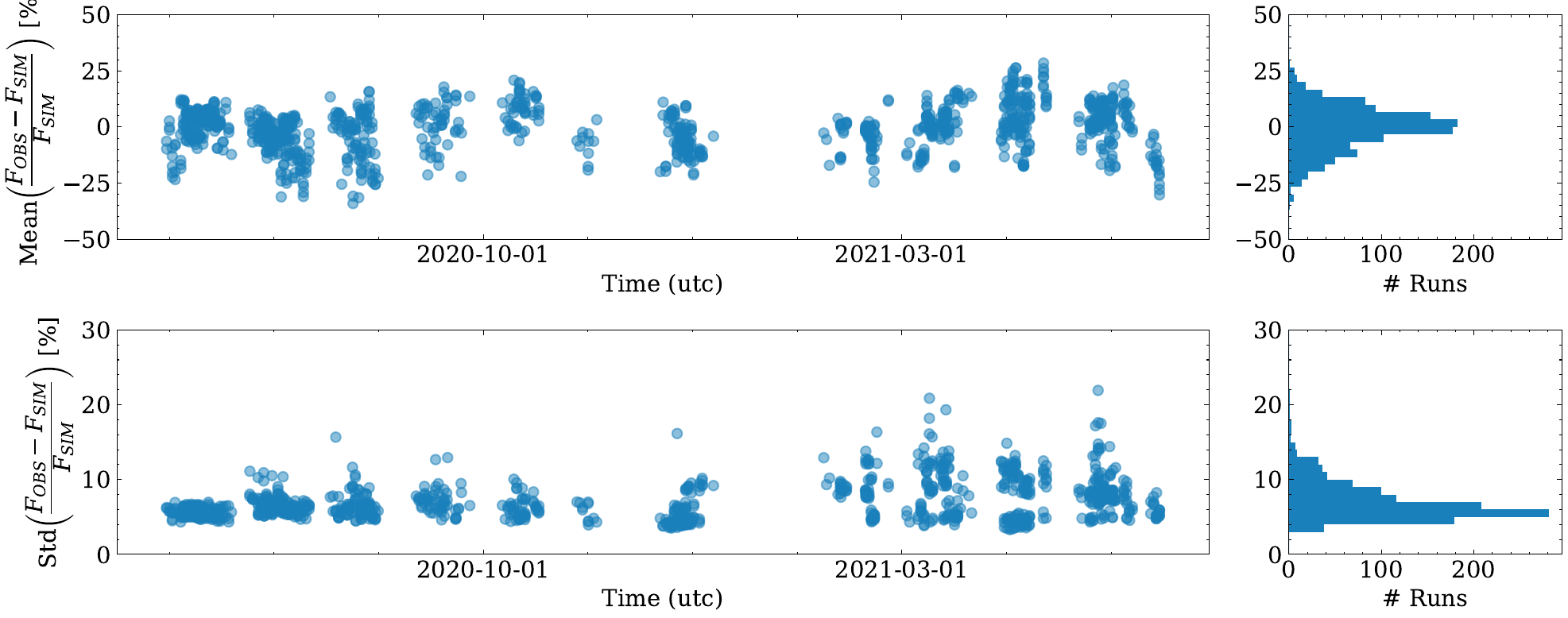}
     \caption{Distribution of the mean and standard deviation of the relative residual $\frac{F_{\mathrm{OBS}}-F_{\mathrm{SIM}}}{F_{\mathrm{SIM}}}$ over all pixels per run over the one year of observations (excluding moonlight runs). The spread in the distribution of the mean (the bias) can be partially explained by the variability of the airglow, see Sect. \ref{ch:ag_var}. The median standard deviation is 6.3\%. The concentration of outliers during the latter half of the year is mainly attributable to observations of Eta Carinae region (for which the NSB is more difficult to predict, see Sect. \ref{ch:etacar}).}
     \label{fig:time_plot}
\end{figure*}
\begin{figure}
  \resizebox{\hsize}{!}{\includegraphics{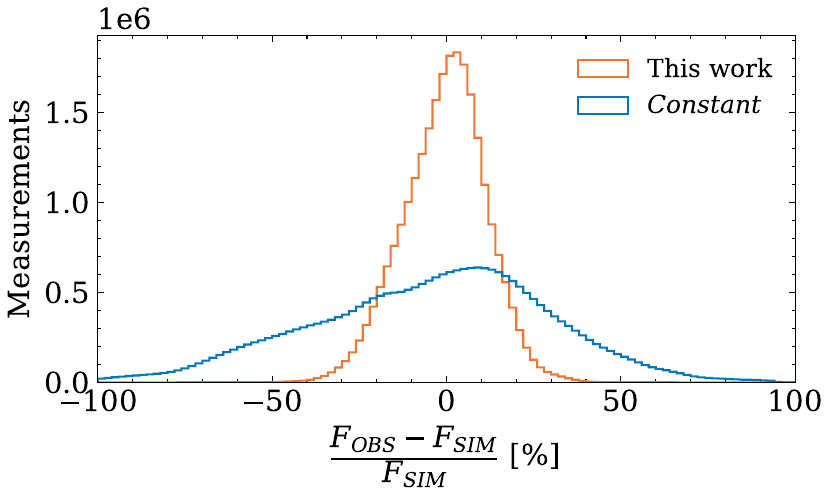}}
  \caption{Distribution of pixel-wise residuals over all pixels for the new method described in this work and the previous method of assuming a constant NSB over the entire FoV of \SI{134}{MHz} for moonless nights.}
  \label{fig:residual}
\end{figure}

Figure \ref{fig:individual_comp} shows the comparison between telescope data (top row) and simulation (middle row) for four different astronomical fields\footnote{For H.E.S.S. gamma-ray source analyses corresponding to these observations, see the following: the Crab Nebula \citep{aharonian_spectrum_2024}, SS 433 \citep{hess_collaboration_acceleration_2024}, the Tarantula Nebula \citep{aharonian_very-high-energy_2024} and M87 \citep{hess_collaboration_curvature_2024}.}. Empty pixels (white) occurring in the top images are inactive pixels or were disabled during the observation because of the presence of a bright object. While the overall shape and inhomogeneity of the NSB data is well reproduced within the simulations, there are deviations from measurements on the order of up to \SI{20}{MHz} for each run. These can most likely be explained by the inability of the diffuse contribution model to capture temporal variations, especially of the airglow component (see Sect. \ref{ch:ag_var}). The standard deviations of the residuals are on the order of \SIrange{5}{10}{\%}. These values are also representative of the total run selection, with the median standard deviation being \SI{6.3}{\%}. Over the course of the year, the mean and standard deviation of the residuals are stable, see Fig. \ref{fig:time_plot}. Compared to the old approach, where the NSB rate was assumed to be uniform and stable at \SI{134}{MHz} for non-moonlight runs, there is a clear improvement as seen in Fig. \ref{fig:residual} (moonlight observation runs were excluded for this comparison). The central $90\%$ range for our method is $[-21 \%, +19 \%]$, while the $90\%$ range for the old approach is $[-64 \%, +48 \%]$.

\section{Discussion}
After introducing the NSB model and its component and describing its application to H.E.S.S.\ data, we now discuss the limitations of the model in the context of systematics and non-modeled effects (Sect. \ref{ch:systematics}); commenting on the effects of airglow variability, atmospheric conditions, bright nebulae and transient events. We also give a short overview of other possible contributions to NSB. In Sect. \ref{ch:usecases}, we describe possible use cases of our new methodology.
\subsection{Systematics / non-modelled effects}\label{ch:systematics}
\subsubsection{Airglow variability}\label{ch:ag_var}
\begin{figure}
  \resizebox{\hsize}{!}{\includegraphics{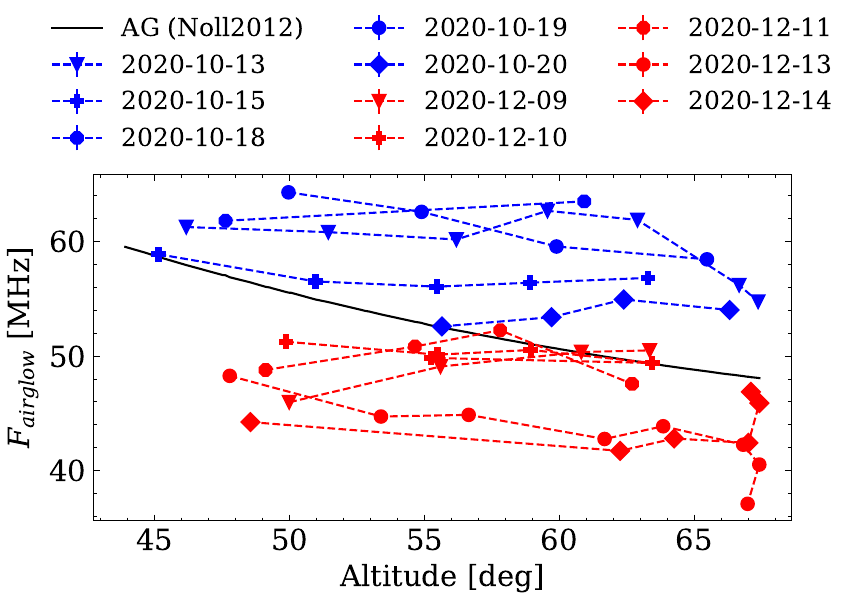}}
  \caption{Airglow fit from Sect. \ref{ch:atm_fit} for 10 nights of observation of an extragalactic field, where the zenith range was larger than \SI{15}{\deg}. For comparison, the predicted value using the model from \cite{noll_atmospheric_2012} is plotted in black.}
  \label{fig:agplot}
\end{figure}
As mentioned in Sect. \ref{ch:airglow}, the airglow has been shown to vary on short (minutes) and long (monthly) timescales during a year \citep{khomich_airglow_2008}. Capturing these variations in a predictive model has proven difficult. 
To quantify the effect of these variations and the accuracy of our predictions, we use the fit to airglow normalisation from Sect. \ref{ch:flatfield} to calculate the airglow strength for H.E.S.S.\ observations of an extragalactic field during the months of October and December of 2020.
We chose this field because the stellar and diffuse galactic contributions are low. It has also been observed multiple times per night and its distance from the celestial pole enables a zenith-dependent analysis. 

As expected, we find strong variability in airglow intensity. Not only is there a difference between the two months, with the average airglow in December being \SI{46.5}{MHz} compared to \SI{58.5}{MHz} in October, there are also night-to-night variations on the order of \SI{10}{MHz}. In some nights, such as 2020-12-13, the airglow can vary on the order of \SI{5}{MHz} within the time span of 1 hour. Given that the median NSB rate for each pixel is around \SI{100}{MHz}, these variations could already be enough to explain the relative prediction error on the order of 12 \% (see Sect. \ref{ch:comparison}.). It also agrees with data taken by \cite{preus_study_2002}, who found the diffuse sky brightness to vary on the order of 10 \% on a nightly basis.

An unexpected discovery is the seeming independence of the airglow on the zenith angle at the H.E.S.S. site. As discussed in Sect. \ref{ch:airglow}, a positive correlation of airglow intensity with zenith angle is expected due to the van Rhijn effect \citep{van_rhijn_brightness_1921}. Comparing the measured airglow values with the model taken from \cite{noll_atmospheric_2012}, there seems to be very little correlation. Interestingly, this is also consistent with observations by \cite{preus_study_2002}, who also found the diffuse sky brightness of dark regions on the sky to be functionally independent of zenith angle at the H.E.S.S.\ site. This merits further study.
\subsubsection{Atmospheric conditions}\label{ch:atm_fit}
\begin{figure}
  \resizebox{\hsize}{!}{\includegraphics{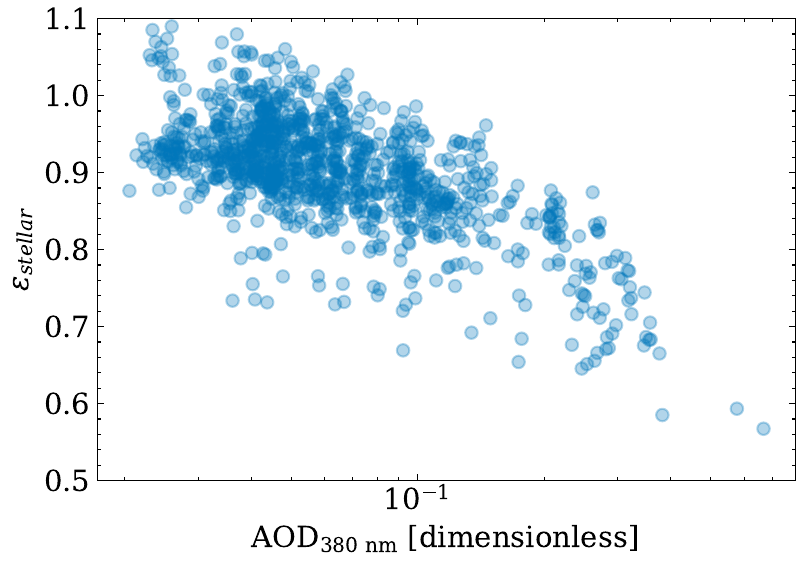}}
  \caption{Correlation between interpolated aerosol optical depth (AOD) at \SI{380}{nm}, as determined during the day by the AERONET system at the H.E.S.S. site, and the normalization of the stellar contribution $\epsilon_{stellar}$ as determined in (\ref{ch:atm_fit}). The Pearson correlation coefficient is $R=-0.63$.}
  \label{fig:aod}
\end{figure}
Atmospheric conditions also have a large effect on the accuracy of the predictions. While the extinction/scattering effect due to Rayleigh scattering changes little, differences in aerosol concentration can lead to large variations in the contribution of stellar and lunar light to the NSB rate in IACTs, while having less of an effect on diffuse sources such as airglow and zodiacal light due to their lower effective scattering coefficient (see Sect. \ref{ch:scattering}). Comparing the fitted stellar normalisation from Sect. \ref{ch:flatfield} and daily aerosol optical depth (AOD) data from the AERONET site at the H.E.S.S.\ observatory, there is an observable correlation with Pearson correlation coefficient $R=-0.63$ (see Fig. \ref{fig:aod}). Since the AOD data is taken during the day, we apply a linear interpolation of the AOD values to estimate aerosol conditions during the night. While this provides a time-resolved approximation, reliance on daytime measurements may not fully capture potentially non-linear variations in AOD throughout the night. As such, deviations between predicted and actual conditions are expected, and a perfect correlation is not expected.
\subsubsection{Bright nebulae} \label{ch:etacar}
\begin{figure}
  \resizebox{\hsize}{!}{\includegraphics{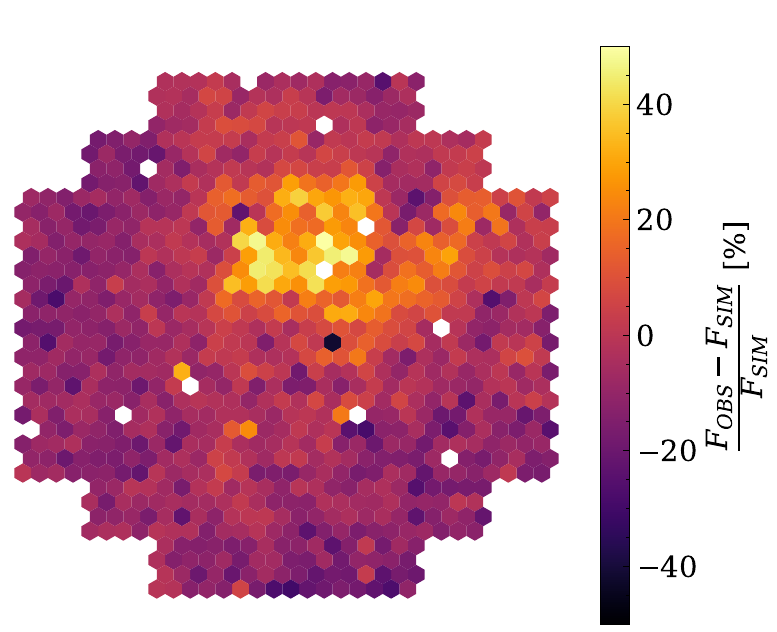}}
  \caption{Residual map of an observation of Eta Carinae. Due to the bright nebula component, the model fails to capture up to 50 \% of the NSB emission coming from the Carina Nebula. For a H.E.S.S. gamma-ray source analysis corresponding to this observation see \cite{aharonian_detection_2025}}
  \label{fig:etacar}
\end{figure}
Due to its construction, the DGL model mainly captures the effect of strong forward scattering as expected in optically thick regions of interstellar dust (see Sect. \ref{ch:galactic}). It is thus unable to capture the effect of bright diffuse emission by optically thick regions. As such, emission and reflection by nebulae is not captured by our model. For bright nebulae, this leads to a measurable underestimation of the flux. In the specific case of the Carina Nebula, which is brightest nebula visible in the night sky and a common target for the H.E.S.S.\ IACTs (see \cite{HESS_collaboration_HESS_2012, HESS_collaboration_detection_2020, aharonian_detection_2025}), this leads to the model to be unable to capture up to 40\% of the emission in the brightest regions of the nebula (see Fig. \ref{fig:etacar}). Since the Carina Nebula is more than an order of magnitude brighter than other nebulae, we expect the effect to be less pronounced for them.
\subsubsection{Transient events and other time-dependent phenomena}\label{ch:transients}
\begin{figure}
  \resizebox{\hsize}{!}{\includegraphics{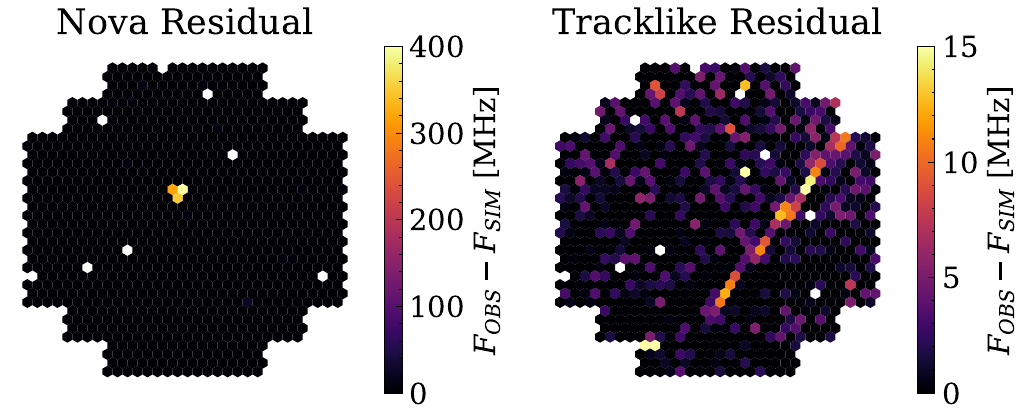}}
  \caption{Residuals caused by a nova and a track-like residual in the dataset, possibly due to a satellite \citep{lang_impact_2023}.}
  \label{fig:nova}
\end{figure}
While the model is able to capture most static components, transients such as satellites, meteors, novae, supernovae and variable stars are not included. Currently, bright satellite or meteor trails affect a sub-percentage of observation time in H.E.S.S. \citep{lang_impact_2023}, but with the emergence of large satellite constellations, this fraction is expected to rise continuously. They could possibly be modelled by using their two-line element sets and albedo, but since two-line element files describing satellite orbits become outdated quickly, these predictions would become inaccurate if done over time periods longer than a few days \citep{bassa_analytical_2022}. Meteor trails cannot be modelled due to their unpredictability. Another example of transient events are bright novae, such as RS Ophiuchi \citep{hess_collaboration_time-resolved_2022}. Such novae can be seen as bright residuals (see Fig. \ref{fig:nova}) on the order of hundreds of MHz.. Variable stars are another issue, since their brightness can change compared to its value in stellar catalogues. For example, over the period of H.E.S.S.\ observations from 2004 to 2024, the star Eta Carinae has gone through significant changes in brightness \citep{martin_changing_2021}. Modelling such stars correctly would require frequent catalogue updates or predictive modelling.
\subsubsection{Other effects}
There is a variety of other effects that could further affect the accuracy of this method. For example, it does not model planets or account for population centres. The closest population centre to the H.E.S.S. site in Namibia is Windhoek, which has a measurable effect in NSB measurements below an altitude of \SI{55}{\degree} \citep{preus_study_2002}. For high-zenith observations, there is also the effect of direct illumination of the pixels due to ground reflections.
\subsection{Use for IACT Analyses}\label{ch:usecases}
\subsubsection{Instrument design}
The ability to make per-pixel estimates of NSB in IACTs is useful for the design of future IACT projects, such as CTAO \citep{the_cherenkov_telescope_array_consortium_science_2019}. One possible area of application is the thermal design of IACT cameras, as high rates of NSB increase the thermal load of silicon photomultipliers, better informing required cooling capacity and thermal safety margins \citep{spencer_advanced_2023}. In the context of instrument passbands, our model can be used to simulate the effect of wavelength filters and PMT choice on expected background rates for different astronomical fields. It is also applicable for the validation of future IACT calibration methods, such as in \cite{abe_star_2023}, where the possibility of online pointing calibration using PMT currents was investigated.
\subsubsection{Observation planning and instrument monitoring}
Accurate simulations of the night sky are beneficial for the planning of IACT observations, especially under moonlight conditions. There are two main considerations: Minimizing the amount of pixels that have to be turned off due to bright stars by choosing the correct field to image a source (since IACTs image most objects at an offset, the choice of field has a rotational freedom), and choosing fields during moonlit nights that reduce the total brightness in the detector (e.g. imaging low-brightness extragalactic targets during moonlit nights). Knowing the brightness in advance can inform the trigger threshold chosen for an observation, avoiding unstable trigger behaviour. It can also help to reduce the amount of turned-off pixels due to high NSB, by lowering the camera gain value based on the astronomical field \citep{bi_performance_2022} Accurate simulations also allow for the comparison of pixel currents or pixel pedestal values with simulated values to identify anomalies, while also enabling the monitoring of optical system degradation, atmospheric conditions (see Sect. \ref{ch:atm_fit}), satellite trails (see Sect. \ref{ch:transients}), and PSF evolution directly from telescope data over time.
\subsubsection{Event classification and reconstruction}
It is possible to augment conventional sim\_telarray simulated images (without NSB) with predictions from our model. A variety of observational conditions could be simulated and be used to create a more realistic training data set for BDT analyses constructed for long-duration use in H.E.S.S..

A future work will discuss how to apply the newly modelled NSB maps to a full IACT analysis, from event images to source spectra. Our NSB tool is best suited for so-called run-wise analyses, whereby simulations are run to correspond to every 28 minutes of observing with H.E.S.S.\ \citep{holler_run-wise_2020}. In this scenario, our tool would be used to generate a map of the average NSB experienced during this time period, and would then be summed with specialised simulations where no other NSB is simulated by packages such as sim\_telarray \citep{bernlohr_simulation_2008}. Alternatively, one could think of a more conventional binned analysis, whereby simulations are generated to represent an average 2D NSB structure in a particular zenith bin.

DL classifiers are known to be particularly sensitive to NSB conditions \citep{shilon_application_2019}. A similar data augmentation approach as in the first paragraph could be used to reduce the discrepancy between simulated training data and IACT observations for supervised machine learning. \cite{dellaiera_deep_2024} showed that tuning simulations with accurate NSB values for an observation leads to improved results comparable or exceeding more complicated domain adaptation algorithms.

\section{Conclusion}\label{ch:conclusion}
In this work, we presented a new method to estimate the night sky background (NSB) in imaging atmospheric Cherenkov telescopes (IACTs) on a per-pixel basis, inspired by techniques used to model the night sky in optical astronomy. We included the diffuse contributions of airglow, zodiacal light, scattered moonlight and diffuse galactic light, relying on models also used for optical astronomy and light pollution modelling \citep{noll_atmospheric_2012, jones_advanced_2013, masana_multi-band_2021, winkler_revised_2022}. We also used GaiaDR3 \citep{gaia_collaboration_gaia_2023} and Hipparcos-Tycho \cite{esa_hipparcos_1997} stellar data, relating it to synthetic stellar spectra from \cite{coelho_new_2014} to calculate the expected contribution to per-pixel NSB using ray-traced pixel effective areas. We accounted for atmospheric extinction and scattering using a single-scattering approximation, simplifying with an effective scattering coefficient when appropriate. 

Comparing our simulation to data from the CT1 telescope of H.E.S.S., we find good agreement between simulation and data, with the central 90\% range of the relative pixel-wise deviation for our method being [-21\%, +19\%], compared to assuming a constant rate, which results in a central 90\% range of [-64\%, +48\%]. We find that there can be deviations on the order of 10\% to 20\%, which could potentially be explained by the variation in atmospheric conditions and the variability of the airglow.

We believe this to be the most sophisticated and accurate simulation of NSB for IACTs to date. A variety of ways it could prove useful to the IACT community are discussed in Sect. \ref{ch:usecases}, including, but not limited to, observation planning and monitoring, instrument design, and studying the effect of NSB on event classification and reconstruction. 

As discussed in Sect. \ref{ch:telescope}, the tool and models used in this work will be released publicly. We believe this to be of significant use to the community. We encourage other experiments such as MAGIC \citep{albert_vhe_2008}, VERITAS \citep{holder_status_2008}, MACE \citep{borwankar_observations_2024}, F.A.C.T. \citep{biland_calibration_2014} and the upcoming CTAO \citep{the_cherenkov_telescope_array_consortium_science_2019}, LACT \citep{zhang_layout_2024} and ASTRI Mini-Array \citep{vercellone_astri_2022} to follow our approach in future modelling of NSB. Our tool can easily be adapted to such instruments, given basic knowledge of, for example, the PDE curve of the photomultipliers used in the camera.

Future studies could expand on the use cases mentioned in Sect. \ref{ch:usecases}, and focus on using time-dependent airglow and atmospheric modelling to reduce the main uncertainties of our model further. They could also expand on the non-modelled effects discussed in Sect. \ref{ch:systematics}, such as adding ground reflections, planets and light pollution models.

\begin{acknowledgements}
This work has been through a review by the H.E.S.S.\ collaboration, who we thank for allowing us to use low level H.E.S.S.\ data in this work, and for useful discussions with collaboration members regarding this work (particularly Alison Mitchell, Simon Steinma{\ss}l and the Innsbruck group). We also thank the other authors of \citet{spencer_advanced_2023} (particularly Rich White, who also provided feedback on this work) and Matthias B{\"u}chele for their previous research on this subject. We also thank Olivier Hainaut for useful discussions. STS is supported by the Deutsche Forschungsgemeinschaft (DFG, German Research Foundation) – Project Number 452934793. The telescope visualisations were done with ctapipe \citep{kosack_cta-observatoryctapipe_2024, cta_ctapipe_2024}. This work made use of Astropy:\footnote{http://www.astropy.org} a community-developed core Python package and an ecosystem of tools and resources for astronomy \citep{astropy_collaboration_astropy_2013,astropy_collaboration_astropy_2018,astropy_collaboration_astropy_2022}. The authors furthermore thank the AERONET project and in particular, Kaleb Negussie as the principal investigator for their effort in establishing and maintaining the HESS AERONET site.
\end{acknowledgements}
\bibliographystyle{aa}
\bibliography{bibliography.bib}
\end{document}